\documentclass[aps,pra,showpacs,twocolumn]{revtex4}
\usepackage{amssymb}
\usepackage{epsfig}
\usepackage{graphicx}
\usepackage{amsmath}
\usepackage{array,color}
\usepackage{multirow}
\usepackage{booktabs}
\usepackage{threeparttable}

\begin{document}

\title{Drag force on a moving impurity in a spin-orbit coupled Bose-Einstein condensate}
\author{Pei-Song He$^{1,2}$, Yao-Hui Zhu$^{1}$ and Wu-Ming Liu$^{2}$}
\address{
$^{1}$School of Science, Beijing Technology and Business University, Beijing 100048, China\\
$^{2}$Beijing National Laboratory for Condensed Matter Physics,
Institute of Physics, Chinese Academy of Sciences, Beijing 100190,
China
}

\date{\today}

\begin{abstract}
We investigate the drag force on a moving impurity in a spin-orbit coupled Bose-Einstein condensate.
We prove rigorously that the superfluid critical velocity is zero when the impurity moves in all but one directions, in contrast to the case of liquid helium and superconductor where it is finite in all directions.
We also find that when the impurity moves in all directions except two special ones, the drag force has nonzero transverse component at small velocity.
When the velocity becomes large and the states of the upper band are also excited, the transverse force becomes very small due to opposite contributions of the two bands.
The characteristics of the superfluid critical velocity and the transverse force are results of the order by disorder mechanism in spin-orbit coupled boson systems.
\end{abstract}

\maketitle

\section{Introduction}
Spin-orbit coupling (SOC) plays a crucial role in many physical systems ranging from nuclei and atoms to quantum spin Hall effect and topological insulators \cite{Bernevig, Zutic, XLQi, Kane}.
An artificial external nonabelian gauge field coupled to neutral atoms of different hyperfine states can be engineered by controlling atom-light interactions \cite{Dudarev, Ruseckas, Juzeli, Stanescu_PRL_2007}.
Recently, Bose-Einstein condensates (BECs) with SOC as well as spin-orbit coupled degenerate Fermi gases have been realized experimentally \cite{YJLin, JYZhang, PWang, Cheuk, Williams}.
It provides physicists with a new platform to study the effects of SOC in many-body systems.
A plenty of researches have been done on the properties of the BEC with SOC, including the ground-state phase \cite{Stanescu_PRA, CWu, Santos, Gopalakrishnan, Ozawa_pra, CWang}, fluctuations above the ground state \cite{Ozawa_pra_2011, KZhou, HZhai_pra_2011, LZhang}, and spin-orbit coupled BECs with other cold-atom techniques, such as dipole-dipole interactions, optical lattice and rotating trap \cite{XQXu, XFZhou, Radic, YDeng, Levenstein}.

One peculiar phenomenon intimately related with BEC is superfluidity, which was successfully explained by Landau \cite{Landau}.
According to the theory, there exists a critical velocity $v_c$ of finite value for an impurity moving in a superfluid, beyond which the impurity experiences a drag force.
This Landau criterion has been confirmed in experiments using ions in superfluid $^{4}\mathrm{He}$ \cite{Ellis} or an optical spoon in a gaseous Bose-Einstein condensate \cite{Raman}.

Landau's analysis can not be directly applied to the case of spin-orbit coupled BEC, since it requires the system to be invariant under Galilean transformation, which is not satisfied when SOC exists.
The low energy excitations of spin-orbit coupled BEC are anisotropic Goldstone modes and spin-waves, both are softer than the Goldstone modes with phonon dispersion in liquid helium.
It is interesting to find out the superfluid critical velocity in the case of a spin-orbit coupled BEC.
Besides, the spin-orbit coupled BEC is anisotropic, and the eigen wave functions of a free boson system have definite helicity, which are opposite for states in the two bands.
It is interesting to investigate whether the moving impurity will experience transverse force, especially for a point-like impurity.
Since the upper band plays role only when the impurity moves fast enough, the effects due to two-band structure on the drag force are also need to be clarified.
In this article, we investigate the superfluidity of a spin-orbit coupled BEC through its effects on an impurity moves in it.
We calculate analytically the drag force and the superfluid critical velocity of the condensate.

This paper is organized as follows:
Sec. \ref{analytical_calculations} gives the model for motion of an impurity in spin-orbit coupled BEC.  We use a time-dependent Gross-Pitaevskii equation to calculate the drag force experienced by the impurity.
In Sec. \ref{result}, the superfluid critical velocity and the drag force are given in details.
Sec. \ref{conclusion} is a summary of this work.
\section{motion of an impurity in spin-orbit coupled BEC}
\label{analytical_calculations}
\begin{figure}[t]
\centering
\includegraphics[width=6cm]{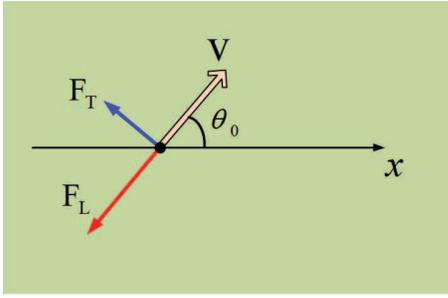}
\caption{ (Color online)
A point-like impurity moves with velocity $\mathbf{v}$ in a two-dimensional spin-orbit coupled Bose-Einstein condensate.
The $x$ direction is the principal direction of the plane-wave condensate.
$\mathrm{F_{L}}$ and $\mathrm{F_{T}}$ are the longitudinal and transverse components of the drag forces experienced by the impurity.
}
\label{BEC}
\end{figure}

We consider a point-like impurity moving in a two-dimensional Rashba spin-orbit coupled BEC with plane-wave order at zero temperature, as shown in Fig. \ref{BEC}.
In the figure, $\mathrm{F_{L}}$ and $\mathrm{F_{T}}$ are longitudinal and transverse components of the drag force experienced by the impurity.
One of the possible realizations of this scenario could be the scattering of heavy neutral molecules by the condensate.

\subsection{An impurity moves in a spin-orbit coupled BEC}

We use a $\delta$-function potential to describe the interaction between the point-like impurity and the bosons in the condensate.
The Hamiltonian for an impurity of $\delta$-function potential moving with constant velocity $\mathbf{v}$ in the BEC is written as
\begin{eqnarray}
\hat{\mathrm{H}} &=& \int d^2\mathbf{r}\hat{\Psi}^+(\mathbf{r},t)[-\frac{\hbar^2}{2m}\nabla^2-\mu-2i\hbar\lambda\nabla\cdot\mathbf{\sigma}]\hat{\Psi}(\mathbf{r},t) \nonumber\\
&+& \int d^2\mathbf{r}\delta(\mathbf{r}-\mathbf{v}t)[g_{i\uparrow}\hat{n}_{\uparrow}(\mathbf{r},t)
+g_{i\downarrow}\hat{n}_{\downarrow}(\mathbf{r},t)] \nonumber\\
&+& \frac{1}{2}\int d^2\mathbf{r}[g_{\uparrow\uparrow}\hat{n}_{\uparrow}(\mathbf{r})^2 + 2g_{\uparrow\downarrow}\hat{n}_{\uparrow}(\mathbf{r})\hat{n}_{\downarrow}(\mathbf{r})
+g_{\downarrow\downarrow}\hat{n}_{\downarrow}(\mathbf{r})^2]. \nonumber\\
\vspace{-15pt}
\label{Hamiltonian}
\end{eqnarray}

In Eq. (\ref{Hamiltonian}), $\hat{\Psi}(\mathbf{r},t) = \left[\hat{\psi}_{\uparrow}(\mathbf{r},t), \hat{\psi}_{\downarrow}(\mathbf{r},t)\right]^{T}$ are time-dependent two-component boson field operators \cite{YJLin}.  $m$ is the mass of the atoms.  $\lambda$ is the strength of SOC.  $\mu$ is the chemical potential.

The momenta of the ground states of the non-interacting boson system of Hamiltonian in Eq. (\ref{Hamiltonian}) compose a ring with $|\mathbf{k}| = \lambda$ in momentum space.
We consider in this work the symmetric point of the two-particle interactions, that is, $g_{\uparrow\uparrow} = g_{\uparrow\downarrow} = g_{\downarrow\downarrow}$.
In this case, quantum fluctuations induced by the particle-particle interactions spontaneously select one momentum out of the ring as the ground state as a result of the order by disorder mechanism \cite{CWang, Villain, Barnett}.
The resulting condensate has plane-wave order.
The condensate is stable against quantum fluctuations in both two and three dimensions \cite{Barnett, Ozawa_PRL_2012, Peisong2013, QZhou}.
It is straightforward to extend our calculations in this paper to the case with $g_{\uparrow\uparrow}g_{\downarrow\downarrow}>g_{\uparrow\downarrow}^2$, in which the condensate also has plane-wave order.
However, the case with $g_{\uparrow\uparrow}g_{\downarrow\downarrow}<g_{\uparrow\downarrow}^2$ is more complicated \cite{QZhou, YLi}, and it will be left to future considerations.

$g_{i\uparrow}$ ($g_{i\downarrow}$) is the strength of the interactions between the impurity and bosons with pseudospin $\uparrow$ ($\downarrow$).
Only $g_{i\uparrow} = g_{i\downarrow}$ will be considered in this work.

The Hamiltonian in Eq. (\ref{Hamiltonian}) has neglected the size effects of the impurity.
Experimentally, it requires the dimension of the impurity to be much smaller than the coherence length $\xi$ of the condensate.
In other cases with impurity of large size, it is natural that the impurity will experience nonzero transverse drag force when it moves in an anisotropic fluid.
The latter is because the properties of the fluid on different points of the impurity's surface are not the same, and this usually leads to a nonzero net transverse force \cite{nematic_liquid_1, nematic_liquid_2, nematic_liquid_3}.
Besides, impurity with large size will also induce vortices from the condensate \cite{Winiecki}.
In spin-orbit coupled BEC, the vortices are generally different from those in case without SOC \cite{Sinha, HHu, Ramachandhran, Kawakami, SWSu, Aftalion}.
These complexities of the size effects of the impurity will be left to future considerations.

\subsection{Drag forces and the Landau criterion of a spin-orbit coupled BEC}
\label{F_vc}

For BEC with plane-wave order, the bosons condensed on one point of the ring with $|\mathbf{k}| = \lambda$ in momentum space, which we set as $\mathbf{k}_0 = (-\lambda, 0)$.
Since the system lacks Galilean invariance due to the existence of SOC \cite{Messiah}, we need to do our calculations of the drag force and the superfluid critical velocity in the frame reference of the static condensate \cite{QZhu}.
In this frame reference, the kinetic momentum of a state with canonical momentum $\mathbf{k}$ is $\mathbf{k}-\mathbf{k}_0$, and the one for the condensate is zero.
In Fig. \ref{BEC}, the positive $x$ direction denotes the principal direction of the plane-wave condensate.

We assume the quantum fluctuations induced by the particle-particle interactions are small.
This is usual in ultracold atom experiments.
We also assume the interactions between the impurity and the bosons are weak.
Then the evolution of the boson fields $\Psi(\mathbf{r},t)$ can be described by the time-dependent Gross-Pitaevskii (GP) equation
\begin{equation}
i\hbar \partial_t \Psi = \left[-\nabla^2-\mu - 2i\lambda\nabla\cdot\mathbf{\sigma} +g_i\delta(\mathbf{r}-\mathbf{v}t)+g|\Psi|^2\right]\Psi.
\label{GP}
\end{equation}

Neglecting the possibility of vortex excitations for point-like impurity, the boson fields including fluctuations induced by the particle-particle interactions and the impurity potential are written as \cite{Peisong2012, WZheng}
\begin{equation}
\Psi(\mathbf{r},t) = \sqrt{\rho_0+\delta\rho} e^{-i\lambda x+i\delta\theta}
\begin{bmatrix}
\cos(\frac{\pi}{4}+\delta\phi) e^{-i\frac{\delta\xi}{2}}\\
\sin(\frac{\pi}{4}+\delta\phi) e^{i\frac{\delta\xi}{2}}
\end{bmatrix},
\label{boson_field}
\end{equation}
where $\rho_0$ is the condensate density, and $\delta\rho(\mathbf{r},t), \delta\theta(\mathbf{r},t), \delta\phi(\mathbf{r},t), \delta\xi(\mathbf{r},t)$ are space-time dependent fluctuations over the condensate wave function.

We substitute the boson fields in Eq. (\ref{boson_field}) into the GP equation in Eq. (\ref{GP}), and then expand the equation up to linear order of the fluctuations.  The result is
\begin{widetext}
\begin{equation}
\partial_{t}
\begin{bmatrix}
\delta\rho(\mathbf{r},t)\\
\delta\theta(\mathbf{r},t)\\
\delta\phi(\mathbf{r},t)\\
\delta\xi(\mathbf{r},t)
\end{bmatrix}
=
\begin{bmatrix}
0&-2\rho_0\nabla^2&0&-2\lambda\rho_0\partial_y\\
-(-\nabla^2+m_{\nu})/(2\rho_0)&0&2\lambda\partial_y&0\\
0&2\lambda\partial_y&4\lambda\partial_x&(-\nabla^2+4\lambda^2)/2\\
-2\lambda\partial_y/\rho_0&0&-(-2\nabla^2+8\lambda^2)&4\lambda\partial_x
\end{bmatrix}
\begin{bmatrix}
\delta\rho(\mathbf{r},t)\\
\delta\theta(\mathbf{r},t)\\
\delta\phi(\mathbf{r},t)\\
\delta\xi(\mathbf{r},t)
\end{bmatrix}
- g_i\delta(\mathbf{r}-\mathbf{v}t)
\begin{bmatrix}
0\\
1\\
0\\
0
\end{bmatrix},
\label{Linearequation}
\end{equation}
\vspace{5pt}
\end{widetext}
where $m_{\nu}$ is the mass for $\delta\rho$ fluctuations, and it equals $2g\rho_0$ in the classical limit \cite{Peisong2012}.
From the last term on the right-hand side of Eq. (\ref{Linearequation}), we find that the impurity potential directly affects $\delta\theta(\mathbf{r},t)$, which are fluctuations of the global phase.
If $g_{i\uparrow}\neq g_{i\downarrow}$ is considered, the impurity will also directly induce $\delta\xi(\mathbf{r},t)$, which are fluctuations of the relative phase between the two components of the bosons.
However, we restrict to the case with equal $g_{i\uparrow}$ and $g_{i\downarrow}$ in this paper.

The drag force with which the condensate acts on the impurity is \cite{Pitaevskii}
\begin{eqnarray}
\vspace{5pt}
\mathbf{F}(t) &=& - \int |\Psi(\mathbf{r},t)|^2 \nabla\left[g_i\delta(\mathbf{r}-\mathbf{v}t)\right] d^2\mathbf{r} \nonumber\\
&=& g_i \left[\nabla|\Psi(\mathbf{r},t)|^2\right]|_{\mathbf{r}=\mathbf{v}t} \nonumber\\
&=& g_i \int i\mathbf{k}  \delta\rho(\mathbf{k}, t) e^{i\mathbf{k}\cdot\mathbf{v}t} d^2\mathbf{k}.
\label{Ft}
\vspace{5pt}
\end{eqnarray}
Since the fluctuations are relative to the condensate wave function, from the definition at the beginning of this subsection we can see that the momenta $\mathbf{k}$ in Eq. (\ref{Ft}) are kinetic ones.
The second line in Eq. (\ref{Ft}) shows the drag force is proportional to the density gradient around the impurity.

The density fluctuations $\delta\rho(\mathbf{k}, t)$ in Eq. (\ref{Ft}) can be obtained from the linearized GP equation in Eq. (\ref{Linearequation}).
To solve Eq. (\ref{Linearequation}), we first perform a Fourier transformation to change it into momentum space.
We find the evolving equations satisfied by fluctuations with different momenta are independent of each other, and for specified momentum $\mathbf{k}$, there is
\vspace{5pt}
\begin{equation}
\frac{\mathrm{\partial}\mathbf{X}(\mathbf{k},t)}{\partial t} = \mathbf{A}(\mathbf{k})\mathbf{X}(\mathbf{k},t)+\mathbf{B}(\mathbf{k},t).
\label{differential_equation}
\end{equation}
Here, $\mathbf{X}(\mathbf{k},t) \equiv [\delta\rho(\mathbf{k},t),
\delta\theta(\mathbf{k},t),
\delta\phi(\mathbf{k},t),
\delta\xi(\mathbf{k},t)]^{T}$.
$\mathbf{A}(\mathbf{k})$ is a $4\times4$ matrix, which is just the one in Eq. (\ref{Linearequation}) rewritten in momentum space,
$\mathbf{B}(\mathbf{k}, t)$ is a four-component vector obtained from Fourier transform of the last term in Eq. (\ref{Linearequation}).
The differential equation (\ref{differential_equation}) can be exactly solved as
\begin{equation}
\mathbf{X}(\mathbf{k}, t) = e^{(t-t_0)\mathbf{A}(\mathbf{k})}\mathbf{X}(\mathbf{k}, t_0) + \int^{t}_{t_0} e^{(t-s)\mathbf{A}(\mathbf{k})} \mathbf{B}(\mathbf{k}, s)ds.
\label{Xt}
\end{equation}
The matrix exponential $e^{t\mathbf{A}}$ in Eq. (\ref{Xt}) can be obtained using the method in Ref. \cite{matrix_exponential}, see Appendix A for details.

The first term on the right-hand side of Eq. (\ref{Xt}) represents the evolution of the quantum fluctuations without the influence of the impurity potential.
It gives zero contributions to the drag force in Eq. (\ref{Ft}).
The second term on the right-hand side of Eq. (\ref{Xt}) represents the evolution of the fluctuations induced by the impurity.
It is proportional to $g_i$, and then from Eq. (\ref{Ft}) we find the drag force depends quadratically on $g_i$.
This is the same to the counterpart in the case without SOC \cite{Pitaevskii}.
We set $t_0 = -\infty$ in Eq. (\ref{Xt}) so as to turn on the impurity potential adiabatically.
In this way, the system will be in a steady state.

Substituting the density fluctuations $\delta\rho(\mathbf{k}, t)$ calculated from Eq. (\ref{Xt}) into the drag force in Eq. (\ref{Ft}), and using analytical continuum to treat the Landau causality \cite{Pitaevskii, causality}, we finally obtain
\begin{widetext}
\begin{eqnarray}
\mathbf{F}
&=&
- 4\pi\rho_0 g_i^2 \int d^2\mathbf{k}\mathbf{k} \left\{\frac{-D+(\omega^{+}_{\mathbf{k}}\omega^{+}_{-\mathbf{k}}+\omega^{+}_{\mathbf{k}}\omega^{-}_{-\mathbf{k}}-\omega^{+}_{-\mathbf{k}}\omega^{-}_{-\mathbf{k}})k^2}{(\omega^{+}_{-\mathbf{k}}+\omega^{-}_{\mathbf{k}})(\omega^{+}_{\mathbf{k}}-\omega^{-}_{\mathbf{k}})(\omega^{-}_{\mathbf{k}}+\omega^{-}_{-\mathbf{k}})}\delta\left(\omega^{-}_{\mathbf{k}}-\mathbf{k}\cdot\mathbf{v}\right) \right.\nonumber\\
&& +\left.\frac{D-(\omega^{+}_{-\mathbf{k}}\omega^{-}_{\mathbf{k}}+\omega^{-}_{\mathbf{k}}\omega^{-}_{-\mathbf{k}}-\omega^{+}_{-\mathbf{k}}\omega^{-}_{-\mathbf{k}})k^2}{(\omega^{+}_{\mathbf{k}}+\omega^{+}_{-\mathbf{k}})(\omega^{+}_{\mathbf{k}}+\omega^{-}_{-\mathbf{k}})(\omega^{+}_{\mathbf{k}}-\omega^{-}_{\mathbf{k}})}\delta\left(\omega^{+}_{\mathbf{k}}-\mathbf{k}\cdot\mathbf{v}\right) \right\},
\label{drag_force}
\end{eqnarray}
\end{widetext}
where $D=k^6 + 16 \lambda^4 k_y^2 + 12 \lambda^2 k^2 k_y^2 + m_{\nu} k^4$.
$\omega^{+}_{\mathbf{k}}$ and $\omega^{-}_{\mathbf{k}}$ are eigenenergies of excitations with momentum $\mathbf{k}$ in the upper and lower bands, respectively \cite{Peisong2012}.
For momenta $\mathbf{k}$ satisfying $\omega^{+}_{\mathbf{k}} = \omega^{-}_{\mathbf{k}}$, the two coefficients before the two delta functions in Eq. (\ref{drag_force}) go zero, and hence the contributions to drag force also vanish.
So the poles in Eq. (\ref{drag_force}) are actually fake.

We also find that $\delta\rho(\mathbf{k}, t)$ obtained from Eq. (\ref{Xt}) has a factor $e^{-i\mathbf{k}\cdot\mathbf{v}t}$, and this factor cancels the term $e^{i\mathbf{k}\cdot\mathbf{v}t}$ in Eq. (\ref{Ft}), and results in the time independence of the drag force in Eq. (\ref{drag_force}).
It is consistent with the fact that the system is in a steady state.
Besides, the density fluctuations induced by the impurity can be written in the form $\delta\rho(\mathbf{r}-\mathbf{v}t)$ in real space.
This form has been taken as an assumption in Ref. \cite{Pitaevskii} to calculate the drag force in the case without SOC.

Using the method above, we can also obtain the drag force for the case without SOC as
$\mathbf{F} =  -2\pi \rho_0 g_i^2 \int d^2\mathbf{k}\mathbf{k} \frac{k^2}{\omega_{\mathbf{k}}}\delta(\omega_{\mathbf{k}}-\mathbf{k}\cdot\mathbf{v}) $.
It is the same to that in Ref. \cite{Pitaevskii}.

The two delta functions in Eq. (\ref{drag_force}) demonstrate that the excitations contributed to the drag force satisfy
\begin{equation}
\omega^{+}_{\mathbf{k}}-\mathbf{k}\cdot\mathbf{v} = 0, \text{or}\ \ \omega^{-}_{\mathbf{k}}-\mathbf{k}\cdot\mathbf{v} = 0.
\label{causality}
\end{equation}
This result is demanded by the causality.
It is the criterion for turning on nonzero drag force in our system.
The form in Eq. (\ref{causality}) is the same to the Landau criterion in Galilean-invariant boson system when the impurity moves in a superfluid at rest \cite{Landau, Ueda}.
Nevertheless, in Galilean-invariant systems, the Landau criterion for a superfluid moving past a resting impurity is $\omega_{\mathbf{k}} + \mathbf{k}\cdot\mathbf{v} = 0$ \cite{Landau, Ueda}.
The two forms of criterions turn into the same one when the system has inversion symmetry $\omega_{\mathbf{k}} = \omega_{-\mathbf{k}}$.
In our system, there is no such symmetry due to the existence of SOC.
It is interesting to study when a superfluid with SOC moves past a resting impurity, whether the second form of the Landau criterion (that is $\omega_{\mathbf{k}} + \mathbf{k}\cdot\mathbf{v} = 0$) still applies.
This will be left to future calculations.

Now we will solve Eq. (\ref{causality}) to obtain the momenta $\mathbf{k}$ of excitations created by the motion of the impurity.
The dispersions of the BEC system, $\omega_{\mathbf{k}}$, satisfy \cite{Peisong2012, Peisong2013}
\begin{equation}
\omega_{\mathbf{k}}^4+b\omega_{\mathbf{k}}^3+c\omega_{\mathbf{k}}^2+d\omega_{\mathbf{k}}+e=0,
\label{condition2}
\end{equation}
where the coefficients are
$b = 8\lambda k_x,
c = -[16\lambda^4+8\lambda^2 k^2-16\lambda^2 k_x^2+8\lambda^2 k_y^2+2k^4+m_{\nu}k^2],
d = -8\lambda k_x[4\lambda^2k_y^2+(k^2+m_{\nu})k^2],
e = (k^4 - 4\lambda^2 k_x^2)^2+m_{\nu}[k_x^2(k^2-4\lambda^2)^2+k^2 k_y^2(k^2+4\lambda^2)]
$.
From Eq. (\ref{condition2}), we obtain that the desired momenta $\mathbf{k}$ for Eq. (\ref{causality}) must satisfy a necessary condition
\begin{equation}
(\mathbf{k}\cdot\mathbf{v})^4+b(\mathbf{k}\cdot\mathbf{v})^3+c(\mathbf{k}\cdot\mathbf{v})^2+d\mathbf{k}\cdot\mathbf{v}+e=0.
\label{condition3}
\end{equation}
We set $k_x = k \cos\theta, k_y = k\sin\theta$, and $v_x = v\cos\theta_0, v_y = v\sin\theta_0$ in polar coordinate system, where $\theta=0$ is taken as the principle direction of the plane-wave condensate, and $\theta_0$ is the direction of the motion of the impurity, see Fig. \ref{BEC}.
By substituting these into Eq. (\ref{condition3}), we obtain
\begin{equation}
k^6+s_4k^4+s_2k^2+s_0=0,
\label{k}
\end{equation}
where
\begin{eqnarray}
s_4 &=& -2[2\lambda\cos\theta+v\cos(\theta-\theta_0)]^2+m_{\nu},\nonumber\\
s_2 &=& [2\lambda\cos\theta+v\cos(\theta-\theta_0)]^4\nonumber\\
    && - 16\lambda^2 v\cos(\theta-\theta_0)[2\lambda\cos\theta+v\cos(\theta-\theta_0)]\nonumber\\
    && + m_{\nu}\{-2[2\lambda\cos\theta+v\cos(\theta-\theta_0)]\nonumber\\
    &&\cdot[6\lambda\cos\theta+v\cos(\theta-\theta_0)]+4\lambda^2\},\nonumber\\
s_0 &=& -16\lambda^4 v^2 \cos^2(\theta-\theta_0) + 16 m_{\nu}\lambda^4 \cos^2\theta.
\label{coefficient}
\end{eqnarray}
When $v$, $\theta$ and $\theta_0$ are specified, Eq. (\ref{k}) is a cubic equation for $k^2$, and hence it can be solved analytically.

The demanded momenta $\mathbf{k}$ are obtained by solving Eq. (\ref{k}) with additional constraints that $k>0$ and $\mathbf{k}\cdot\mathbf{v}\ge0$.
Whether the obtained $\mathbf{k}$ belongs to excitations of the upper or lower band will be checked by Eq. (\ref{causality}).
A short prove on the validity of the above method to solve Eq. (\ref{causality}) is given in Appendix B.

In polar coordinates, the integration $\int d^2\mathbf{k}$ in Eq. (\ref{drag_force}) is changed into $\int d\theta \int dk k$.
The two delta functions in Eq. (\ref{drag_force}) are removed analytically by first employing the formula \cite{Appel}
\begin{equation}
\delta(f(k)) = \sum_{i}\frac{1}{|f'(k)|_{k=k_i}}\delta(k-k_i),
\label{deltafunction}
\end{equation}
and then doing an integration over $k$.
To calculate the denominator in Eq. (\ref{deltafunction}), we need the expressions that
\begin{equation}
\frac{\partial\omega^{\pm}_{\mathbf{k}}}{\partial k} = -\frac{\frac{\partial b}{\partial k}(\omega^{\pm}_{\mathbf{k}})^3+\frac{\partial c}{\partial k}(\omega^{\pm}_{\mathbf{k}})^2+\frac{\partial d}{\partial k}\omega^{\pm}_{\mathbf{k}}+\frac{\partial e}{\partial k}}{4(\omega^{\pm}_{\mathbf{k}})^3+3b(\omega^{\pm}_{\mathbf{k}})^2+2c\omega^{\pm}_{\mathbf{k}}+d}.
\label{domega_dk}
\end{equation}
The above equation is obtained from the fact that Eq. (\ref{condition2}) are satisfied for all momentum $\mathbf{k}$,
and so when the left-hand side is differentiated by $\mathbf{k}$, it still gives zero.
The resulting expressions lead to Eq. (\ref{domega_dk}).

\section{Drag force and superfluid critical velocity}
\label{result}

In this section, we present the drag force $\mathbf{F}$ experienced by the moving impurity in the spin-orbit coupled BEC.
The superfluid critical velocity $v_c$ is then obtained.

\begin{figure}[t]
\centering
\includegraphics[width=8.5cm]{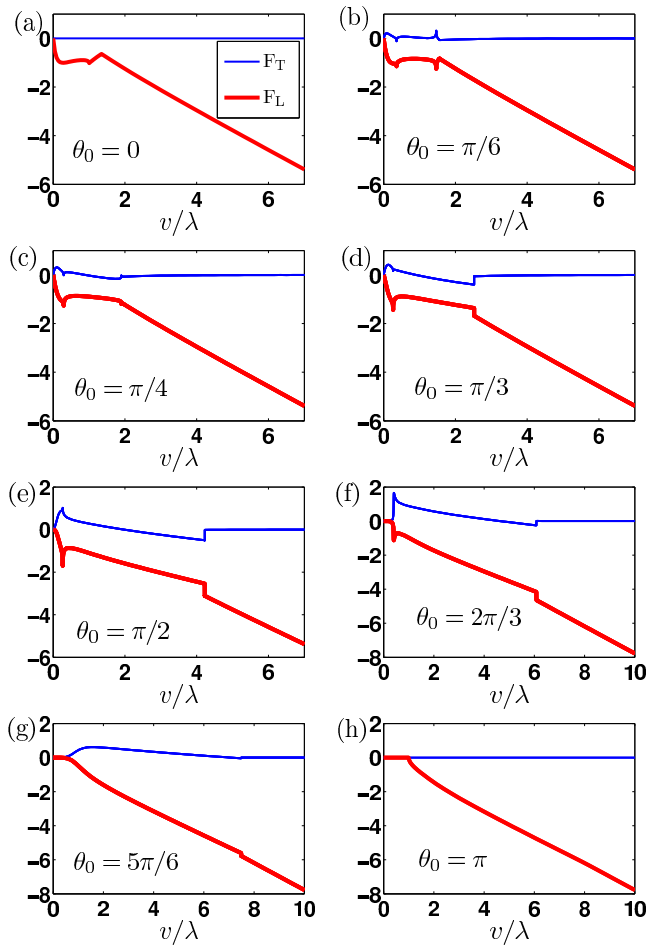}
\caption{(Color online)
The $v/\lambda$-dependence of the longitudinal and transverse drag forces experienced by the impurity.
$\theta_0$ increases from (a) to (h).
$\tilde{m}_{\nu} \equiv m_{\nu}/\lambda^2 = 1$ is used in the calculations.
$\mathrm{F_{L}}$, $\mathrm{F_{T}}$ are in unit of $4\pi\rho_0 g_i^2 \lambda$.
The characteristics of the drag force as $\theta_0$ varies are summarized in Table \ref{table:classification}.
}
\label{FL_FT}
\end{figure}

In Fig. \ref{FL_FT}, we present the drag forces as a function of $v/\lambda$ for some typical values of $\theta_0$.
We use $\mathrm{F_{L}}$ and $\mathrm{F_{T}}$ to denote the longitudinal and transverse components of the drag force, respectively, and there is
\begin{eqnarray}
\mathrm{F_{L}} &\equiv& \mathrm{F_x} \cos\theta_0 +\mathrm{F_y} \sin\theta_0, \nonumber\\
\mathrm{F_{T}} &\equiv& -\mathrm{F_x} \sin\theta_0 +\mathrm{F_y} \cos\theta_0.
\label{Fxy_to_FLT}
\end{eqnarray}
Since the boson system (not including the impurity) is symmetric with respect to the $k_x$ axis, we only need to consider the cases with $0\le\theta_0\le\pi$.
The definite values of the drag forces depend on the value of $\tilde{m}_{\nu}\equiv m_{\nu}/\lambda^2$.
However, in situations with weak quantum fluctuations, the results are qualitatively the same.

\begin{table}[h]
\renewcommand{\arraystretch}{1.8}
\begin{threeparttable}
\caption{ Classifications of the $v/\lambda$-dependence of the drag force according to $\theta_0$}
\vspace{5pt}
\centering 
\begin{tabular}{c|ccccc} \toprule[2px]
                          &\multicolumn{5}{c}{\multirow{2}{*}{$\theta_0$}}\\
                          &\multicolumn{5}{c}{}                                                  \\\hline
\multirow{2}{*}{$v_c$}    &\multicolumn{4}{c}{$[0, \pi)$}             &$\{\pi\}$\ \  \\
                          &\multicolumn{4}{c}{$=0$}                     &$>0$                     \\\hline
\ \ existence of \ \ \ \      &\multicolumn{2}{c}{$\{0\}$}          &\multicolumn{2}{c}{$(0, \pi)$}         &$\{\pi\}$\ \   \\
nonzero $\mathrm{F_{T}}$         &\multicolumn{2}{c}{No}  &\multicolumn{2}{c}{Yes}   &No       \\\hline
$\ \ \ \mathbf{F} \approx 0$ at\ \ \ \  &\multicolumn{2}{c}{$[0, \pi/2]$}          &\multicolumn{2}{c}{$(\pi/2, \pi)$}         &$\{\pi\}$\ \   \\
small $v/\lambda$         &\multicolumn{2}{c}{No}  &\multicolumn{2}{c}{Yes}   &$\mathrm{F} =0$       \\\hline
\multirow{2}{*}{existence of}     &\multicolumn{2}{c}{$\ \ \ [0, 0.27\pi)^*\ $} &\multicolumn{2}{c}{$\ \ \ [0.27\pi, \pi)^*\ \ $}  &$\{\pi\}$\ \   \\
a jump                          &\multicolumn{2}{c}{No}    &\multicolumn{2}{c}{Yes}     &No                 \\\hline
\multirow{2}{*}{existence of}     &\multicolumn{3}{c}{$\ \ \ \ \ \ \ \ \ \ \ \ \ \ [0, 0.65\pi]^*\ \ \ \ $}        &\multicolumn{2}{c}{$\ \ \ \ \ \ \ \ (0.65\pi, \pi]^*$}\ \ \      \\
peaks                          &\multicolumn{3}{c}{\ \ \ \ \ \ \ \ Yes}    &\multicolumn{2}{c}{\ \ \ \ \ \ No} \\\bottomrule[2px]
\end{tabular}
\label{table:classification}
\vspace{5pt}
 \begin{tablenotes}
 \item[*] The values of the bounds $0.27\pi, 0.65\pi$ depend on the value $m_{\nu}/\lambda^2$, as shown in Fig. \ref{bound}, and here we take $m_{\nu}/\lambda^2$=1 in our calculations.
 \end{tablenotes}
\end{threeparttable}
\end{table}

\begin{figure}[t]
\centering
\includegraphics[width=8.5cm]{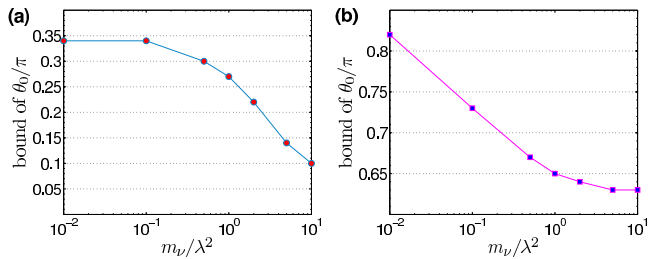}
\caption{ (Color online)
The $m_{\nu}/\lambda^2$-dependence of the bounds that appear in the last two rows of the Table \ref{table:classification} for (a) existence of a jump in the drag force; (b) existence of peaks in the drag force.
}
\label{bound}
\end{figure}

In Table \ref{table:classification}, we have summarized the characteristics of the $v/\lambda$-dependence of the drag forces according to the value of $\theta_0$.
In the second row of the table, it states that the superfluid critical velocity $v_c$ is nonzero only for $\theta_0 = \pi$,
which has zero weight in the phase space of $\theta_0$.
In the third row of the table, it says that there exist nonzero transverse force for motion in directions $0<\theta_0<\pi$.
The fourth row of the table states that the drag force is very tiny at small $v/\lambda$ for motion in directions $\pi/2<\theta_0<\pi$, while for motion in other directions, the drag force is considerable at small $v/\lambda$.
The fifth row shows that there is a jump in the value of the drag force as $v/\lambda$ varies for $0.27\pi\le\theta_0<\pi$,
where the lower bound $\theta_0=0.27\pi$ depends on the value of $\tilde{m}_{\nu}$.
The $\tilde{m}_{\nu}$-dependence of this bound is shown in Fig. \ref{bound}(a) in the region $10^{-2}\le \tilde{m}_{\nu} \le 10$, which is typical in ultracold atom experiments.
The last row of the table states that for $0\le\theta_0<0.65\pi$, there are peaks in drag force as a function of $v/\lambda$.
The peak here means the dependence of the drag force on $v/\lambda$ is nonanalytic.
The upper bound $\theta_0 = 0.65\pi$ also depends on the value of $\tilde{m}_{\nu}$, as shown in Fig. \ref{bound}(b).

Besides the results in Table \ref{table:classification}, there is $\mathrm{F_{T}} \simeq 0$ and $\mathrm{F_{L}}\propto v/\lambda$ (in unit of $4\pi\rho_0 g_i^2\lambda$) at large $v/\lambda$ for all $\theta_0$.

\subsection{Superfluid critical velocity}
\label{subsec_vc}

As listed in the second row of Table \ref{table:classification}, the superfluid critical velocity $v_c$ is nonzero only when the impurity moves in the direction $\theta_0=\pi$.
In the following we will prove this result analytically.

Since Eq. (\ref{k}) is a cubic equation for $k^2$,  it is easy to find that when coefficient $s_0$ is negative, Eq. (\ref{k}) have at least one positive root.
It means that, when the velocity of the impurity is given, then for any $\theta$ satisfies
\begin{equation}
s_0 = -16\lambda^4 v^2 \cos^2(\theta-\theta_0) + 16 m_{\nu}\lambda^4 \cos^2\theta<0,
\label{vc_prove_1}
\end{equation}
Eq. (\ref{k}) will have solutions with $k>0$.
There is an additional constraint $\mathbf{k}\cdot\mathbf{v}\ge0$, which leads to $\cos(\theta-\theta_0)\ge0$.
This constraint together with Eq. (\ref{vc_prove_1}) gives
\begin{equation}
\arctan \frac{\frac{\sqrt{m_{\nu}}}{v}-\cos\theta_0}{\sin\theta_0} < \theta < \pi - \arctan \frac{\frac{\sqrt{m_{\nu}}}{v}+\cos\theta_0}{\sin\theta_0}.
\label{s0_1}
\end{equation}
For convenience, in the above inequality, $\theta$ is defined in the region $[-\pi, \pi)$ when $0<\theta_0\le\pi/2$, and it is defined in the region $[0, 2\pi)$ when $\pi/2<\theta_0<\pi$.
For any $v$ and $m_{\nu}$ of finite value, the inequality (\ref{s0_1}) gives a finite range of $\theta$.
Since any excitation contributes a negative quantity to $\mathrm{F_{L}}$, the total contributions of the excitations satisfy the inequality (\ref{s0_1}) are finite after the momentum integration being taken.
The cause that for any finite $v$ there exist states excited by the impurity as given in Eq. (\ref{s0_1}) lies in the anisotropy of the Goldstone modes,
which are softer than phonon with linear dispersion.

Besides exciting the Goldstone modes, when the impurity moves in directions $0\le\theta_0\le\pi/2$, there are always a group of spin-waves with momenta around $\mathbf{k} = (2\lambda, 0)$ to be excited for any finite $v$.
It is because the eigenvalues of the spin-waves go zero when their momenta go to $\mathbf{k} = (2\lambda, 0)$.

In a word, the superfluid critical velocity $v_c =0$ for $0\le\theta_0<\pi$.  This is one of the main results in this work.
Both the facts that the Goldstone modes have dispersions softer than phonon and the existence of the soft spin-waves are results of spontaneously breaking of the infinitely degenerate ground states of Rashba spin-orbit coupled boson system by quantum fluctuations, which is called order by disorder mechanism \cite{Barnett, Villain}.

\subsection{Impurity valve}

As listed in the fourth row of Table \ref{table:classification}, when $v/\lambda$ is small, there is enormous difference between the transports of the impurity starts from the two ends of the spin-orbit coupled BEC in Fig. \ref{BEC}:  it experiences considerable drag force when moves in direnctions $0<\theta_0\le\pi/2$,
while in directions $\pi/2<\theta_0<\pi$ the drag force is tiny.

\begin{figure}[t]
\centering
\includegraphics[width=8cm]{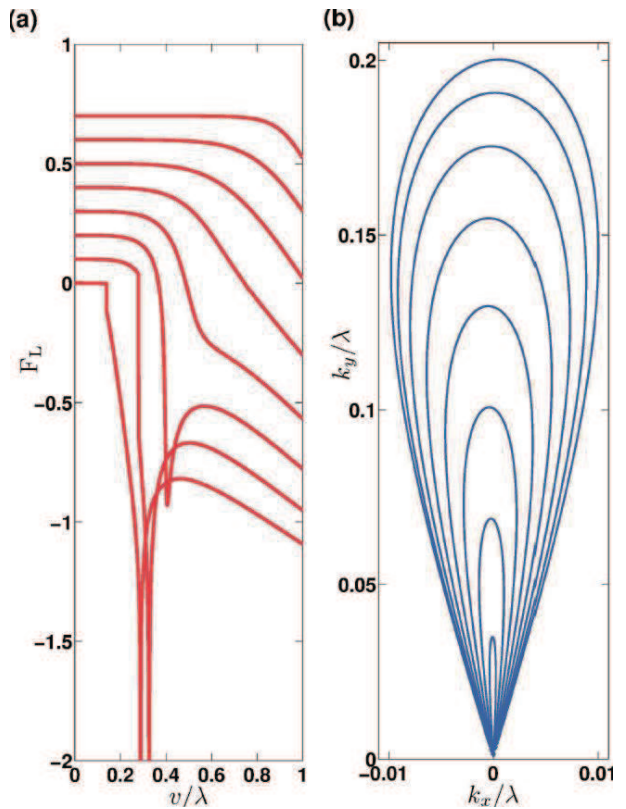}
\caption{ (Color online)
(a) $\mathrm{F_{L}}$ at small $v/\lambda$ for $\theta_0 = \pi/2+j\pi/18$, $j= 1,2, ..., 8$, from left to right.  For clarity, the lines have been shifted upwards in steps as $\theta_0$ increases.  $\mathrm{F_{L}}$ are in unit of $4\pi\rho_0 g_i^2 \lambda$.  $\mathrm{F_{L}}$ is tiny at small $v/\lambda$.
(b) The momenta of the excitations over the condensate due to the motion of the impurity at $v/\lambda = 0.1$ for the $\theta_0$ in (a).  The loops go large as $\theta_0$ increases.  The momenta of these Goldstone modes are quite small, and result in tiny drag forces.
}
\label{tiny_F}
\end{figure}

The difference lies in the exciting of spin-waves at small $v/\lambda$ in directions $0<\theta_0\le\pi/2$.
They give a large contributions to the drag force.
Although there are Goldstone modes excited by the impurity in both cases, their contributions to the drag force are tiny when the velocity of the impurity is small.
This is illustrated in Fig. \ref{tiny_F}.
In Fig. \ref{tiny_F}(a), we show the $\mathrm{F_{L}}$ with $0<v/\lambda<1$ for $\theta_0 = \pi/2 + j\pi/18, j=1, 2, .., 8$.
We take $v/\lambda=0.1$ as an example.
We see the drag forces are quite small for all values of $\theta_0$ at this velocity.
Fig. \ref{tiny_F}(b) shows the momenta of excitations for all the $\theta_0$ in Fig. \ref{tiny_F}(a) at this velocity.
The excitations consist of Goldstone modes only.
We see that the momenta of the excitations are quite small.
Besides, they compose a small phase space in momentum space.
Simple calculations show their contributions to the drag force are tiny.

This property implies that the spin-orbit coupled BEC can be employed as a potential impurity valve.
In applications, the principal direction of the condensate (or equally the momentum of condensate wave function) can be controlled by applying an infinitesimal magnetic field or a shift of the wavelength of the vector light coupled with the hyperfine states of ultracold atoms \cite{Radic, YJLin_2009}.

\vspace{-10pt}
\subsection{Transverse force}

When the velocity of the impurity is not too large, generally the impurity experiences a finite transverse force, unless it moves along the symmetry axis of the condensate, as shown in Fig. \ref{FL_FT}.
Moreover, the transverse force can reverse its direction when $v/\lambda$ varies.
When the velocity of the impurity is large enough, the transverse force goes tiny for all $\theta_0$.
There exists a sudden jump between these two regions when the impurity moves in directions $0.27\pi\le\theta_0<\pi$,
while it evolves continuously for motion in directions $0<\theta_0<0.27\pi$.
The following part of this subsection will dwell on interpreting these properties of the system.

\begin{figure}[t]
\centering
\includegraphics[width=8.5cm]{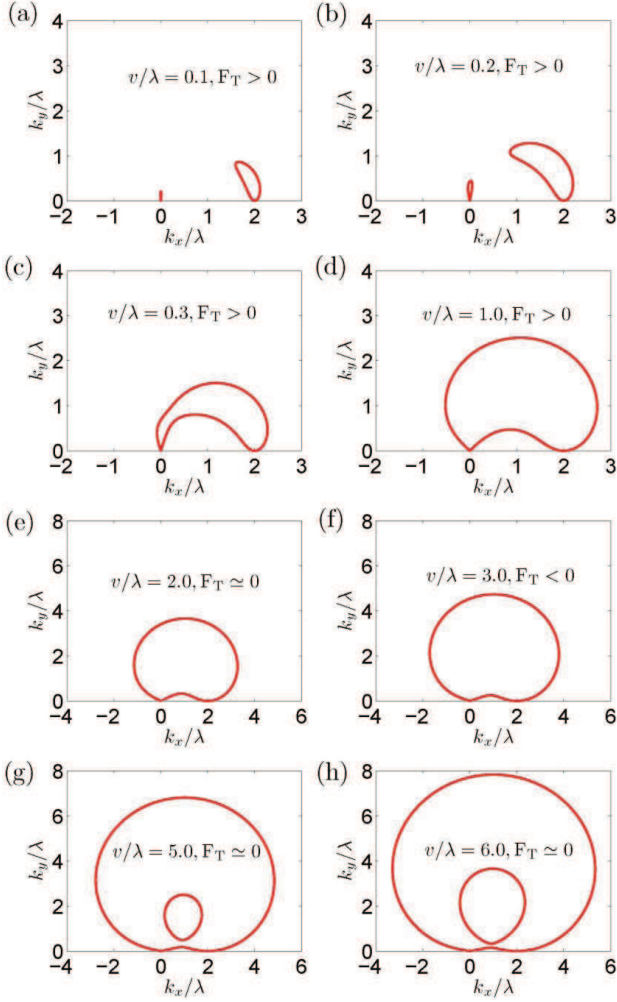}
\caption{ (Color online)
The momenta of excitations for $\theta_0 = \pi/2$ as $v/\lambda$ increases.
From (a) to (h), $v/\lambda = 0.1, 0.2, 0.3, 1.0, 2.0, 3.0, 5.0, 6.0$.
In (a)-(c), the spin-waves around $\mathbf{k}=(2\lambda, 0)$ dominate over the Goldstone modes around $\mathbf{k} = (0, 0)$, and it leads to $\mathrm{F_{T}}>0$.
From (d) to (f), the excitations with large momenta gradually dominate over the spin-waves, and as a result, $\mathrm{F_{T}}$ changes continuously from positive to negative.
In (g) and (h), the upper band (the smaller loops in the figures) also takes part in, and the cancellation with the contributions from the lower band makes  $\mathrm{F_{T}}\simeq0$.
}
\label{kx_ky_FT_90d}
\end{figure}

\subsubsection{Transverse force at small $v/\lambda$}

For convenience, we express the drag force in Eq. (\ref{drag_force}) as $\mathbf{F} = \mathbf{F}^{-} + \mathbf{F}^{+}$, with
\begin{eqnarray}
\mathbf{F}^{-} &\equiv& - 4\pi\rho_0 g_i^2 \int d^2\mathbf{k}\mathbf{k}f^{-}(\mathbf{k})\delta(\omega^{-}_{\mathbf{k}}-\mathbf{k}\cdot\mathbf{v}), \nonumber\\
\mathbf{F}^{+} &\equiv& - 4\pi\rho_0 g_i^2 \int d^2\mathbf{k}\mathbf{k} f^{+}(\mathbf{k})\delta(\omega^{+}_{\mathbf{k}}-\mathbf{k}\cdot\mathbf{v}),
\label{FPM}
\end{eqnarray}
where $f^{\pm}(\mathbf{k})$ denote corresponding factors before the delta functions.

The original Hamiltonian of the system Eq. (\ref{Hamiltonian}) is symmetric with respect to the direction of $\mathbf{v}$.
Naively, it implies zero $\mathrm{F_{T}}$.
However, the O(2) symmetry of the bosons system in momentum space is broken by quantum fluctuations via selecting a single state from the macroscopic denegerate ground states when the Bose-Einstein condensation occurs.
Only the symmetry with $k_y\leftrightarrow -k_y$ is remained for the BEC system.
This has the consequence that in Eq. (\ref{FPM}),
$f^{\pm}(\mathbf{k})$ have only the inversion symmetry $\mathbf{k}\leftrightarrow -\mathbf{k}$,
and $\omega^{\pm}_{\mathbf{k}}$ are only symmetric under $k_y \leftrightarrow -k_y$.
This is different from that of $\mathbf{k}\cdot\mathbf{v}$ in Eq. (\ref{FPM}), which is symmetric with respect to the direction of $\mathbf{v}$.
As a result, the expressions of $\mathbf{F}^{\pm}$ in Eq. (\ref{FPM}) and then $\mathbf{F}$ are not symmetric with respect to the direction of $\mathbf{v}$.
This can be seen more clearly at $\theta_0 = \pi/2$,
in which the transverse force is along the symmetry axis of the condensate.
From Eqs. (\ref{Fxy_to_FLT}) and (\ref{FPM}), we have for $\theta_0 = \pi/2$
\begin{eqnarray}
\mathrm{F_{T}} &=& 4\pi\rho_0 g_i^2 \int^{\infty}_{-\infty}d k_x \int^{\infty}_{0} d k_y k_x[f^{-}(\mathbf{k})\delta(\omega^{-}_{\mathbf{k}}-k_y v) \nonumber\\
&& + f^{+}(\mathbf{k})\delta(\omega^{+}_{\mathbf{k}}-k_y v)].
\label{Eq:FT90d}
\end{eqnarray}
Since $\omega^{\pm}_{\mathbf{k}}$ are not even functions of $k_x$ \cite{Peisong2012}, there is generally $\mathrm{F_{T}} \neq 0$.
In short, merely from the symmetry analysis, we find that the anisotropy of the BEC will generally leads to a nonzero transverse force.

In the followings, we further take $\theta_0 = \pi/2$ as an example to investigate the properties of $\mathrm{F_{T}}$, and also the central physical processes involved.
In Fig. \ref{kx_ky_FT_90d}, we exhibit the momenta of the excited states for various values of $v/\lambda$.

At small $v/\lambda$ as in Figs. \ref{kx_ky_FT_90d}(a)(b), the contour of the momenta is composed of two loops: one consists of Goldstone modes around the condensed momentum, and the other consists of spin-waves around $\mathbf{k} = (2\lambda, 0)$.
Only the states in the lower band are excited in these cases.

It is easy to see that the transverse force is mainly due to spin-waves in these two cases, and the Goldstone modes give negligible contributions.
Firstly, for $\theta_0=\pi/2$, the contribution to the transverse force by an excitations is proportional to $k_x$, and the coefficient is positive since there is $f^{-}(\mathbf{k})\ge0$ for any momentum $\mathbf{k}$.
$k_x$ of the spin-waves are around $2\lambda$, and it is large and always positive.
While $k_x$ of the Goldstone modes are quite small, and the contributions from the excitations with $k_x>0$ and $k_x<0$ will further offset each other.
Secondly,
as shown clearly in Figs. \ref{kx_ky_FT_90d}(a)(b), the phase space of the spin-waves in momentum space is much larger than that of the Goldstone modes.
Summarising above analysis, we conclude that at small $v/\lambda$ as in Figs. \ref{kx_ky_FT_90d}(a)(b), the spin-waves dominate over the Goldstone modes on contributing to the transverse force, and the result is positive.
Besides, it increases as $v/\lambda$ goes up, since a larger amount of spin-waves will be excited as $v/\lambda$ goes larger.
This agrees with the result in Fig. \ref{FL_FT}(e).

\begin{figure}[t]
\centering
\includegraphics[width=8.0cm]{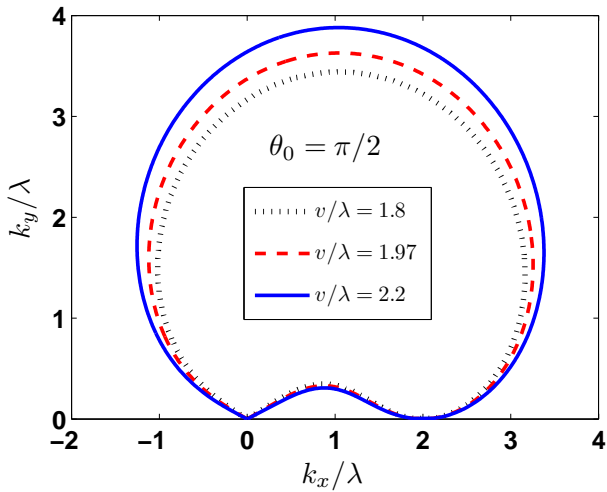}
\caption{ (Color online)
The momenta of excitations for $\theta_0 = \pi/2$ as $v/\lambda$ increases, with the transverse force turning from positive to negative.
There is: $\mathrm{F_{T}}>0$ for $v/\lambda=1.8$; $\mathrm{F_{T}}\simeq0$ for $v/\lambda=1.97$; $\mathrm{F_{T}}<0$ for $v/\lambda=2.2$.
The excitations with large momenta gradually dominate over the spin-waves as $v/\lambda$ increases.
}
\label{change_sign_90d}
\end{figure}

When $v/\lambda$ increases further, the two loops get larger, and then they merge into a single one, as seen in Fig. \ref{kx_ky_FT_90d}(c).
The peak of $\mathrm{F_{T}}$ at $v/\lambda\simeq0.27$ corresponds to the critical point when the two loops merge.
Detailed analysis will be given in the Sec. \ref{section:peak}.

After that, $\mathrm{F_{T}}$ gradually decreases as $v/\lambda$ goes up.
At the same time, the corresponding loop of the contour of the momenta gets larger.
$\mathrm{F_{T}}$ decreases to approximately zero at $v/\lambda$ about 2.0 (1.97 in precision), and then evolves continuously to negative.
We can see that the contours of the momenta in Figs. \ref{kx_ky_FT_90d}(d)-(f) are similar in shapes, except that their sizes are different.

To make it more clear, in Fig. \ref{change_sign_90d}, we compare the contours of the momenta for $v/\lambda = 1.8, 1.97$ and $2.2$, where $\mathrm{F_{T}}\simeq0$ at $v/\lambda=1.97$.
We find that the parts of the three loops with $k_y$ small have little difference.
Their contributions to the transverse forces are almost the same.
Just like the cases in Figs. \ref{change_sign_90d}(a)(b), spin-waves dominate this part, and they provide a positive quantity for $\mathrm{F_{T}}$.
However, there exist distinct differences among the parts of the three loops with large $k_y$.
The energies of the excitations in this part are large.
We will see from Eq. (\ref{Fscaling-2}) in next subsection that the transverse force from contributions of these states is a linear function of $v/\lambda$ with a negative coefficient.
So they contribute a negative quantity to the transverse force, and this quantity decreases as $v/\lambda$ goes from $1.8$ to $2.0$.
By summarising the contributions of the two parts, we obtain that $\mathrm{F_{T}}$ is a decreasing function of $v/\lambda$ for these values of $v/\lambda$, and it is possible that $\mathrm{F_{T}}$ will turn negative when $v/\lambda$ is beyond some value (say $v/\lambda = 1.97$ for $\theta_0 = \pi/2$).

When $v/\lambda$ increases further, as shown in Figs. \ref{kx_ky_FT_90d}(g)(h), the upper band is also turned on, and $\mathrm{F_{T}}$ goes tiny.
It indicates that the two bands give opposite contributions to $\mathrm{F_{T}}$.
The physical origins of this result will be given in the following two parts.

\subsubsection{Drag force at large $v/\lambda$}
\label{scaling}

We find there is $\mathrm{F_{T}}\simeq0$ and $\mathrm{F_{L}}\propto v/\lambda$ at large $v/\lambda$, where the drag forces are in unit of $4\pi\rho_0 g_i^2\lambda$, as shown in Fig. \ref{FL_FT}.
This can be simply interpreted by rewritten the drag force in Eq. (\ref{drag_force}) as
\begin{equation}
\mathbf{F}(\lambda, m_{\nu}, \mathbf{v}) = v\mathbf{\tilde{F}}\left(\frac{\lambda}{v}, \frac{m_{\nu}}{v^2}, \frac{\mathbf{v}}{v}\right),
\end{equation}
where $\mathbf{\tilde{F}}$ is the dimensionless form of $\mathbf{F}$ scaled by $v$.
In the limit of large $v/\lambda$, there is $\lambda/v\simeq0$, and $\mathbf{\tilde{F}}\left(\lambda/v, m_{\nu}/v, \mathbf{v}/v\right)$ has no manifest dependence on the SOC strength $\lambda$.
Then, the drag force $\mathbf{F} = v\mathbf{\tilde{F}}$ behaves like the one for the case without SOC.
In the latter case, $\mathrm{F_{T}} = 0$ and $\mathrm{F_{L}}\propto v$ at large $v$.

\begin{figure}[t]
\centering
\includegraphics[width = 8cm]{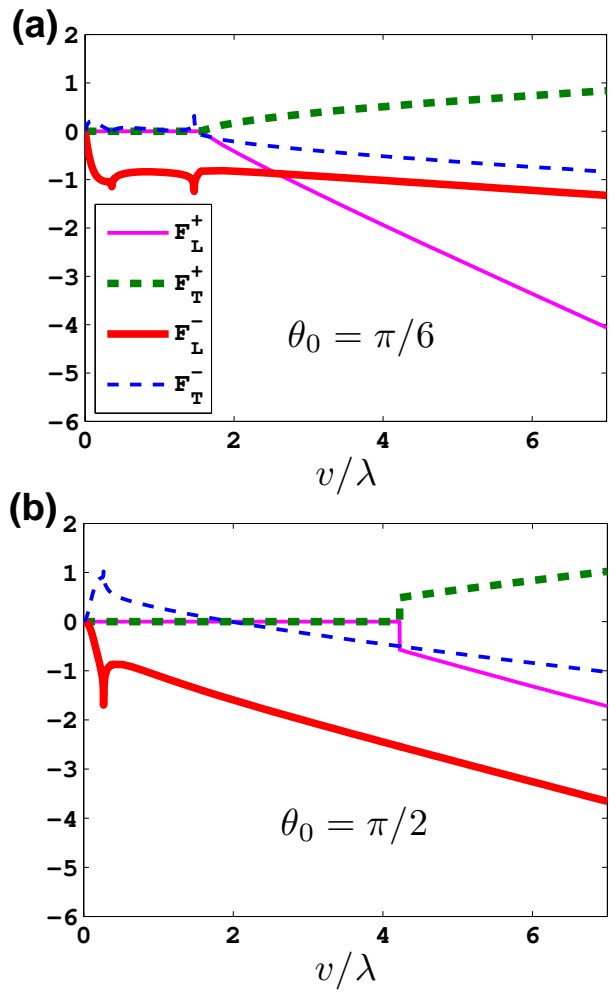}
\caption{(Color online) The drag forces (in unit of $4\pi\rho_0 g_i^2 \lambda$) as a function of $v/\lambda$ for (a) $\theta_0 = \pi/6$ and (b) $\theta_0 = \pi/2$.
In (a) and (b), at large $v/\lambda$, all of the lines are in linear of $v/\lambda$, and $\mathrm{F^{+}_{T}}\simeq -\mathrm{F^{-}_{T}} >0$.
The jumps of $\mathrm{F^{+}_{L}}$ and $\mathrm{F^{+}_{T}}$ in (b) are due to turning on the excitations of the upper band, which give finite contributions at the tangent point between the plane $\mathbf{k}\cdot\mathbf{v}$ and the band of $\omega^{+}_{\mathbf{k}}$, and immediately $\mathrm{F_{T}} = \mathrm{F^{+}_{T}} + \mathrm{F^{-}_{T}} \simeq0$.  There are no jumps in (a) when the upper band starts to play role, since the states around the touch point of the two bands are first excited, which give vanishing contributions.
}
\label{v_FP_FM}
\end{figure}

We also find that in the limit of large $v/\lambda$, $\mathrm{F^{\pm}_{L}}$ and $\mathrm{F^{\pm}_{T}}$ are in linear of $v/\lambda$, in unit of $4\pi\rho_0 g_i^2\lambda$, for the motion of the impurity in all directions $\theta_0$.
Two examples with $\theta_0 = \pi/2$ and $\pi/6$ are shown in Fig. \ref{v_FP_FM}.
The two cases represent typical ones with and without a jump when the states in the upper band just begin to be excited, respectively.
In Fig. \ref{v_FP_FM}, we find $\mathrm{F^{\pm}_{L}}<0$ and $\mathrm{F^{+}_{T}} \simeq - \mathrm{F^{-}_{T}}<0$ at large $v/\lambda$ in both cases.
Moreover, we will find that these two properties are satisfied for all $0<\theta_0<\pi$.

In the followings, we will obtain the expressions for these approximate behaviors by expanding the drag force in Eq. (\ref{drag_force}) in powers of $v/\lambda$ at large $v/\lambda$.

We will first calculate the leading order term of the drag force at large $v/\lambda$, which is in linear of $v/\lambda$.
It is easy to find from the expression of the drag force in Eq. (\ref{drag_force}) that the leading order term is mainly contributed by states with large momenta.
In the following we will consider only excitations with large $\mathbf{k}$.
The contributions of excitations with small momenta will not affect coefficient of the term in linear of $v/\lambda$.
At large $k$, the dispersions of excitations are $\omega^{\pm}_{\mathbf{k}} = \left(\sqrt{(k_x-\lambda)^2+k_y^2}\pm\lambda\right)^2+C^{\pm}(\mathbf{k})m_{\nu}+O(1/k)$, where $C^{\pm}(\mathbf{k})$ are anisotropic with $0\le C^{\pm}(\mathbf{k})\le \frac{1}{2}$ and $C^{+}(\mathbf{k})+C^{-}(\mathbf{k}) = \frac{1}{2}$ \cite{omega_uv}.
By substituting the above expression into the drag force in Eq. (\ref{drag_force}), and then expanding it in powers of $v/\lambda$, we finally obtain the leading order term as
\begin{eqnarray}
\mathbf{F} &\simeq& - 4\pi\rho_0 g_i^2 \cdot \frac{1}{4}\int d^2\mathbf{k} \mathbf{k} \left[\left(1-\frac{k_x}{k}\right)\delta\left(\omega^{-}_{\mathbf{k}}-\mathbf{k}\cdot\mathbf{v}\right)\right.\nonumber\\
&&+\left.\left(1+\frac{k_x}{k}\right)\delta\left(\omega^{+}_{\mathbf{k}}-\mathbf{k}\cdot\mathbf{v}\right)\right].
\label{Fscaling-1}
\end{eqnarray}
The momenta satisfy the delta functions $\delta(\omega^{\pm}_{\mathbf{k}}-\mathbf{k}\cdot\mathbf{v})$ in Eq. (\ref{Fscaling-1}) are also solved in expansion of $v/\lambda$.
In polar coordinates, there is
\begin{equation}
k^{\pm} = \lambda \left[\cos(\theta-\theta_0)v/\lambda+2(\cos\theta\mp1)+O\left(\frac{1}{v/\lambda}\right)\right],
\label{kpm}
\end{equation}
for specified $\theta$.
Here, the fact that $k^{\pm}>0$ restricts the possible values of $\theta$.

We then substitute Eq. (\ref{kpm}) into Eq. (\ref{Fscaling-1}).
We first do the integration over $k$ for specified $\theta$ to remove the delta functions,
and then do the integration over $\theta$.
The longitudinal and transverse components of the drag forces contributed from the upper and lower bands are
\begin{eqnarray}
\mathrm{F^{\pm}_{L}} &=& -4\pi\rho_0 g_i^2\lambda\cdot\left\{\frac{1}{3}\left[\frac{3\pi}{8}\pm\cos\theta_0\right]v/\lambda +O\left(\left(v/\lambda\right)^{0}\right)\right\},\nonumber\\
\mathrm{F^{\pm}_{T}} &=& \pm4\pi\rho_0 g_i^2\lambda\cdot\left\{\frac{\sin\theta_0}{6}v/\lambda + O\left(\left(v/\lambda\right)^{0}\right)\right\},
\label{Fscaling-2}
\end{eqnarray}
respectively, where the $O\left(\left(v/\lambda\right)^{0}\right)$ terms are functions of $m_{\nu}/\lambda^2$, $\mathbf{k}/\lambda$.

The slopes of the linear functions in Eq. (\ref{Fscaling-2}) are in good agreement with the results in Fig. \ref{v_FP_FM}.
They depend on the values of $\theta_0$.
For $0<\theta_0<\pi$, these slopes are always nonzero.
Moreover, the ones for $\mathrm{F^{\pm}_{L}}$ are always negative, while the ones for $\mathrm{F^{+}_{T}}$ and $\mathrm{F^{-}_{T}}$ are opposite to each other.
When $\theta_0=\pi/2$, the slope of $\mathrm{F^{-}_{T}}$ is negative.
It means that the total contributions of the excitations with large momenta to $\mathrm{F^{-}_{T}}$ is negative.
This is consistent with the result in Fig. \ref{kx_ky_FT_90d} that when $v/\lambda$ goes large, the value of the transverse force can be negative.

The next-to-leading terms of the drag forces in Eq. (\ref{Fscaling-2}), which is of order $O\left(\left(v/\lambda\right)^{0}\right)$,  are of finite values, and will shift the lines.

The longitudinal drag force $\mathrm{F_{L}}$, which equals $\mathrm{F^{+}_{L}}+\mathrm{F^{-}_{L}}$, has
\begin{eqnarray}
\mathrm{F_{L}} &=& -4\pi\rho_0 g_i^2\lambda\cdot\left\{\frac{\pi}{4}v/\lambda + O\left(\left(v/\lambda\right)^{0}\right)\right\}.
\label{Fscaling-3}
\end{eqnarray}
The slope does not depend on the direction of motion $\theta_0$.
This agrees with the results in Fig. \ref{FL_FT} at large $v/\lambda$.

In Eq. (\ref{Fscaling-1}), only the terms $\pm\frac{k_x}{k}$ in the factors $1\pm\frac{k_x}{k}$ give contributions to $\mathrm{F^{\pm}_{T}}$ after the momentum integration.
It means the contributions of the excitations from the two bands with the same momentum (when it is large) to the transverse force are opposite to each other.
In the free boson system, the upper and lower bands have opposite helicities \cite{HuiZhai_review}, which is the eigenvalue of the helicity operator $\hat{h} = \mathbf{k}\cdot\mathbf{\sigma}/|\mathbf{k}|$.
In the interacting boson system, the excitations at large momenta behave like free bosons.
It can be easily obtained that exciting a single particle with definite helicity by the moving impurity will induce a nonzero transverse force, whose direction will be reversed if the helicity changes its sign.

Furthermore, from a straightforward but lengthy calculation (see details in Appendix C),
we find the term of order $O\left(\left(v/\lambda\right)^{0}\right)$ for $\mathrm{F_{T}}$, which equals $\mathrm{F^{+}_{T}} + \mathrm{F^{-}_{T}}$, is exactly zero.  That is
\begin{equation}
\mathrm{F_{T}} = - 4\pi\rho_0 g_i^2\lambda\cdot O\left(\frac{1}{v/\lambda}\right).
\label{FT_large_v}
\end{equation}
It is very small at large $v/\lambda$.
Moreover, this result does not depend on the direction of the motion of the impurity.
This is consistent with our results in Fig. \ref{FL_FT}.

\subsubsection{Jump in drag force as $v/\lambda$ varies}

We find there is a finite jump in $\mathrm{F_{L}}$ and also $\mathrm{F_{T}}$ as $v/\lambda$ varies when the impurity moves in directions $0.27\pi\le\theta_0<\pi$, while there is no jump for the motion in other directions.
The height of the jump evolves continuously from a finite value in $0.27\pi\le\theta_0<\pi$ to zero in other directions.
Here, the lower bound $0.27\pi$ depends on the value of $m_{\nu}/\lambda^2$, as shown in Fig. \ref{bound}(a).

For the impurity moving in directions $0.27\pi\le\theta_0<\pi$, there is a threshold velocity $v_{th}$ for specified $\theta_0$, when $v\ge v_{th}$ the excitations in the upper band contribute to the drag force.
At the threshold velocity $v_{th}$, only one state in the upper band is excited.
We denote its momentum as $\mathbf{k}_{th} = k_{th}(\cos\theta_{th}, \sin\theta_{th})$.
It satisfies
\begin{eqnarray}
\frac{\partial\omega^{+}_{\mathbf{k}}}{\partial k}\bigg{|}_{k = k_{th}} - v_{th}\cos(\theta_{th}-\theta_0) &=& 0, \nonumber\\
\frac{\partial\omega^{+}_{\mathbf{k}}}{\partial \theta}\bigg{|}_{\theta = \theta_{th}} + k_{th}v_{th}\sin(\theta_{th}-\theta_0) &=& 0.
\label{upperband-singlek}
\end{eqnarray}
Its contribution to the drag force is
\begin{eqnarray}
\mathbf{F}^{+}_{th}
&=&  - 4\pi\rho_0 g_i^2 \int d^2\mathbf{k}\mathbf{k} f^{+}(\mathbf{k})\delta(\omega^{+}_{\mathbf{k}} - \mathbf{k}\cdot\mathbf{v}_{th})\nonumber\\
&=& - 4\pi\rho_0 g_i^2 \int d^2\mathbf{k} \mathbf{k} f^{+}(\mathbf{k})C\delta(k - k_{th})\delta(\theta-\theta_{th}), \nonumber\\
\label{Eq_jump}
\end{eqnarray}
where $C$ is a finite number proportional to the curvature of function $\omega^{+}_{\mathbf{k}}-\mathbf{k}\cdot\mathbf{v}_{th}$ in momentum space.
Since $\mathbf{k}_{th}f^{+}(\mathbf{k}_{th})$ is generally nonzero, $\mathbf{F}^{+}_{th}$ in Eq. (\ref{Eq_jump}) gives a finite contribution to the drag force.
As a result, the drag force displays a jump at $v_{th}$.

Furthermore, it is easy to prove that for $0.27\pi\le\theta_0<\pi$, the $\theta_{th}$ obtained from Eq. (\ref{upperband-singlek}) has $\theta_{th}-\theta_0\neq0$.
That is, the momentum of the excitation from the upper band at $v_{th}$ is not on the direction of $\theta_0$.
So its contribution to the drag force has transverse component, see $\theta_0 = \pi/2$ in Fig. \ref{v_FP_FM}(b) as an example.

The case with $\theta_0 = \pi$ is an exception.
Although there exists a solution satisfies Eq. (\ref{upperband-singlek}), it makes $f^{+}(\mathbf{k}_{th})=0$.
So there is no jump when the velocity of the impurity reaches $v_{th}$.

The situations with $0\le\theta_0<0.27\pi$ are different from the above cases.
For motion in these directions, the state where the upper and lower bands touch is excited at $v = v_{th}$.
The coefficient $f^{+}(\mathbf{k})$ is zero at the momentum of this state.
So there is no jump of $\mathbf{F}$ at $v_{th}$, see $\theta_0 = \pi/6$ in Fig. \ref{v_FP_FM}(a) as an example.

\begin{figure}[t]
\includegraphics[width=8.5cm]{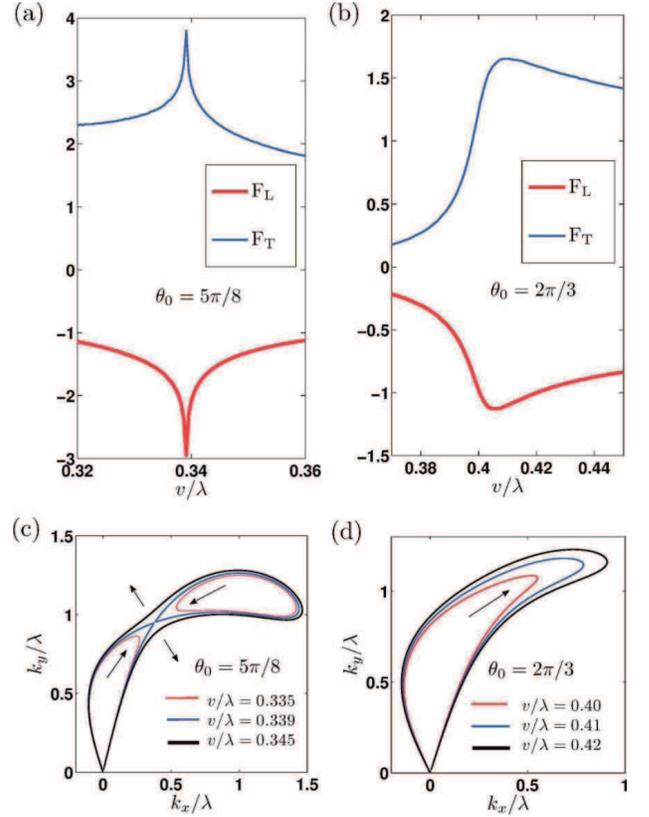}
\vspace{5pt}
\caption{(Color online)
(a): The $v/\lambda$-dependence of drag forces have nonanalytic peaks.
(b): The local maxima of the drag forces is an analytic hump.
(c): The contour of momenta of excitations changes topologically from two loops to one loop when $v/\lambda$ varies across the peak in (a).  The peak corresponds to the critical point when two loops merge into one.  The arrows indicate the evolution direction of the loops as $v/\lambda$ increases.
(d): The contour of the momenta of excitations evolves without topological change when $v/\lambda$ varies across the local minimum of the hump in (b).
}
\label{peak_vs_hump}
\end{figure}

\subsection{Peaks in drag force as a function of $v/\lambda$}
\label{section:peak}

We find for $0\le\theta_0\le0.65\pi$, there exist peaks in the drag forces as a function of $v/\lambda$,
where the upper bound $0.65\pi$ depends on the value of $m_{\nu}/\lambda^2$, as shown in Fig. \ref{bound}(b).
This nonanalytic behavior is due to the topological changes in the contour of the momenta of the excitations as $v/\lambda$ varies.
The peak corresponds to the critical point when the number of the loops changes, see $\theta_0 = 5\pi/8$ in Fig. \ref{peak_vs_hump}(a) as an example.

But for $0.65<\theta_0\le\pi$, there exist only humps, in which the drag forces are analytic functions of $v/\lambda$.
The contour of the momenta of the excitations evolves without topological changes when the local maximum (or minimum) of the drag forces is passed through as $v/\lambda$ varies.

Since the drag force is proportional to the gradient of the density of states in real space, see the second line in Eq. (\ref{Ft}), the above results mean the density of states is a nonanalytic function of $v/\lambda$ when peaks exist, while it is an analytic one when only humps exist.

\section{Conclusions}
\label{conclusion}

In conclusion, we have considered a point-like impurity moves with constant velocity in a two-dimensional spin-orbit coupled Bose-Einstein condensate.
Base on a time-dependent Gross-Pitaevskii equation, we calculate analytically the drag force experienced by the impurity.
Besides, we have proved rigorously that the superfluid critical velocity is zero for motion of impurity in all but one direction.
This is because in these directions there is always exciting of anisotropy Goldstone modes which have dispersions softer than phonon.
We also find that there exists enormous difference for the impurity to be scattered into the Bose-Einstein condensate from two opposite ends.
It is due to the scattering of spin-waves by the impurity, which exists only for the impurity to move from one end of the condensate.
The anisotropic Goldstone modes with dispersions softer than phonon and also the existence of the spin-waves are both resulting from the mechanism of order by disorder in a Rashba spin-orbit coupled boson system.

We also find that there is nonzero transverse force when the impurity moves with not large velocity in all directions except two special ones.
The transverse force is due to the anisotropy of the Bose-Einstein condensate, and the spin-waves play the crucial role in it.  When the impurity moves fast, the transverse force is extremely small due to cancellation of the contributions from states in the upper and lower bands, respectively.  This is because when the impurity moves with a large velocity, the contributions to the transverse force are dominated by states with large momenta, which have opposite helicities for the two bands.
The direction of the contributed transverse force will be reversed if the helicity of the excitation changes its sign.
Our results will be of help to make it more clear the superfluidity of the spin-orbit coupled Bose-Einstein condensate.

Experimentally, our results can be verified by scattering a heavy neutral molecule into the BEC cloud and detecting its track and velocity.
The longitudinal drag force will slow down the motion of the impurity.
The transverse force makes the track of impurity curving, and the sign of the force reflects as the direction of the curvature.
We assume the spin-orbit coupled $^{87}\mathrm{Rb}$ Bose-Einstein condensate is confined in a harmonic trap with oscillator frequencies $(f_x, f_y, f_z) = (50\mathrm{Hz}, 50\mathrm{Hz}, 1000\mathrm{Hz})$.
The size of the condensate in-plane is about $4\mu$m.
The density of the condensate is modulated to $2.4\times10^{10}\mathrm{cm}^{-2}$, which corresponds to $m_{\nu}/\lambda^2\simeq1$.
We consider a molecule with mass about ten times of that for $^{87}\mathrm{Rb}$ atom, and the $s$-wave scattering length between the molecule and $^{87}\mathrm{Rb}$ atom about the same as that between two $^{87}\mathrm{Rb}$ atoms.
When it is scattered into the condensate in the direction $\theta_0=\pi/2$ with velocity $v=1\mathrm{mm/s}$, which gives $v/\lambda=0.5$, then according to our calculations, the transverse deflection is about 40nm.
However, to have a delay time up to $10\%$ of that need to pass through the condensate freely, the density of bosons needs to be at least $2.4\times10^{11}\mathrm{cm}^{-2}$.

\section*{ACKNOWLEDGEMENTS}
This work was supported by
the National Key Basic Research Special Foundation of China (NKBRSFC) under grants Nos. 2011CB921502, 2012CB821305,
the National Natural Science Foundation of China (NSFC) under Grants Nos. 61227902, 61378017, 11174020 and 11075176;
Beijing Natural Science Foundation (BNSF) under Grant No. 2102014 and No. 1112007;
the Starting up Foundation for Youth Teachers of Beijing Technology and Business University under Grants No. QNJJ20123-19,
the Project Sponsored by the Scientific Research Foundation for the Returned Overseas Chinese Scholars, State Education Ministry,
Funding for training talents in Beijing City with Project No. 2011D005003000012,
and PXM2013-014213-000013.
\begin{appendix}

\section{matrix exponential $e^{t\mathrm{\mathbf{A}}}$}
\label{appendix_etA}

In this appendix, we will give the details of the calculations of the matrix exponential $e^{t\mathbf{A}}$ in Eq. (\ref{Xt}) and also the drag force in Eq. (\ref{drag_force}).
The matrix exponential $e^{t\mathbf{A}}$ is derived following the method in Ref. \cite{matrix_exponential}.

The four eigenvalues of the matrix $\mathbf{A}$ in Eq. (\ref{differential_equation}) are $\lambda_j = -i\omega^{(j)}_{\mathbf{k}}, j=1,2,3,4$, where $\omega^{(j)}_{\mathbf{k}}$ are four roots of Eq. (\ref{condition2}) for specified momentum $\mathbf{k}$, that is $-\omega^{-}_{-\mathbf{k}}, -\omega^{+}_{-\mathbf{k}}, \omega^{+}_{\mathbf{k}}, \omega^{-}_{\mathbf{k}}$.
The exponential of matrix $\mathbf{A}$ is obtained as
\begin{equation}
e^{t\mathbf{A}} = \varphi_1(t)\mathbf{I} + \varphi_2(t)\mathbf{A} + \varphi_2(t)\mathbf{A}^2 + \varphi_3(t)\mathbf{A}^3,
\label{etA1}
\end{equation}
where \begin{equation}
[\varphi_1(t)\ \varphi_2(t)\ \varphi_3(t)\ \varphi_4(t)] = [y_1(t)\ y_2(t)\ y_3(t)\ y_4(t)]\mathbf{W}[y; 0]^{-1},
\label{etA2}
\end{equation}
and $y_j(t) = e^{\lambda_j t}, j=1,2,3,4$.
$\mathbf{I}$ is the $4\times4$ unit matrix.
$\mathbf{W}[y;t]$ is the Wronski matrix
\begin{equation}
\mathbf{W}[y; t] =
\begin{bmatrix}
y_1(t)&y_2(t)&y_3(t)&y_4(t)\\
y'_1(t)&y'_2(t)&y'_3(t)&y'_4(t)\\
y''_1(t)&y''_2(t)&y''_3(t)&y''_4(t)\\
y'''_1(t)&y'''_2(t)&y'''_3(t)&y'''_4(t)
\end{bmatrix}.
\label{etA3}
\end{equation}

From Eqs. (\ref{etA1})-(\ref{etA3}), the drag force is
\begin{eqnarray}
\mathbf{F}(t)
&=& g_i \int i\mathbf{k}  d^2\mathbf{k} \delta\rho(\mathbf{k}, t) e^{i\mathbf{k}\cdot\mathbf{v}t}\nonumber\\
&=& - 4\pi\rho_0 g_i^2\int d^2\mathbf{k} i\mathbf{k} \int^{\infty}_{0} ds [e^{s\mathrm{A}}]_{12}e^{i\mathbf{k}\cdot\mathbf{v}s},
\end{eqnarray}
where
\begin{eqnarray}
[e^{t\mathbf{A}}]_{1,2}
&=& \sum_{j=1}^{4} y_j(t)\bigg{\{}\mathbf{A}_{1,2}\Big[\mathbf{W}[y; 0]^{-1}\Big]_{j,2}\nonumber\\
&&+[\mathbf{A}^2]_{1,2}\Big[\mathbf{W}[y; 0]^{-1}\Big]_{j,3}\nonumber\\
&&+[\mathbf{A}^3]_{1,2}\Big[\mathbf{W}[y; 0]^{-1}\Big]_{j,4} \bigg{\}}.
\end{eqnarray}
This gives the result in Eq. (\ref{drag_force}).


\section{method to solve the momenta of excitations}

In this appendix, we will give short prove on the validity of the method to solve Eq. (\ref{causality}) in Sec. \ref{F_vc}.

Firstly, for specified momentum $\mathbf{k}$, Eq. (\ref{condition2}) gives four roots $\omega^{\pm}_{\mathbf{k}}, -\omega^{\pm}_{-\mathbf{k}}$.
The detailed expressions as a function of $\mathbf{k}$ are given in Ref. \cite{Peisong2012}.
Secondly, we solve Eq. (\ref{condition3}) with the constraint $k>0$ to obtain the momentum $\mathbf{k}$.
Then, we substitute this momentum into the four expression of $\omega^{\pm}_{\mathbf{k}}, -\omega^{\pm}_{-\mathbf{k}}$.
The results are represented as $\omega^{(i)}_{\mathbf{k}}, (i =1,2,3,4)$, respectively.
Finally, we substitute $\omega^{(i)}_{\mathbf{k}}, (i =1,2,3,4)$ into Eq. $(\ref{condition2})$, and then
subtract Eq. $(\ref{condition3})$.
The result is
\begin{eqnarray}
&&\left(\omega^{(i)}_{\mathbf{k}}-\mathbf{k}\cdot\mathbf{v}\right)\left\{\left(\omega^{(i)}_{\mathbf{k}}+\mathbf{k}\cdot\mathbf{v}\right)\left[\left(\omega^{(i)}_{\mathbf{k}}\right)^2+\left(\mathbf{k}\cdot\mathbf{v}\right)^2\right]\right.\nonumber\\
&&+b\left[\left(\omega^{(i)}_{\mathbf{k}}\right)^2+\omega^{(i)}_{\mathbf{k}}(\mathbf{k}\cdot\mathbf{v})+(\mathbf{k}\cdot\mathbf{v})^2\right]\nonumber\\
&&+c\left(\omega^{(i)}_{\mathbf{k}}+\mathbf{k}\cdot\mathbf{v}\right)+d\bigg\} = 0, (i=1,2,3,4).
\label{condition23}
\end{eqnarray}
Since $\omega^{(i)}_{\mathbf{k}}, (i =1,2,3,4)$ are all different from each other, then for specific $\mathbf{k}$ there exists and only exists one $\omega^{(i)}_{\mathbf{k}}$ which makes $\omega^{(i)}_{\mathbf{k}}-\mathbf{k}\cdot\mathbf{v}$=0.
It equals $\omega^{+}_{\mathbf{k}}$ or $\omega^{-}_{\mathbf{k}}$, since there is $\mathbf{k}\cdot\mathbf{v}\ge0$.

\section{transverse force up to order $O\Big((v/\lambda)^0\Big)$ at large $v/\lambda$}
\label{appendix_ft}

We will give the detailed calculations of transverse force $\mathrm{F_{T}}$ up to order $O\Big((v/\lambda)^0\Big)$ at large $v/\lambda$.
The result is shown in Eq. (\ref{FT_large_v}).

Using Eq. (\ref{Fxy_to_FLT}), the transverse force $\mathrm{F_{T}}$ obtained from Eq. (\ref{drag_force}) is
\begin{eqnarray}
\mathrm{F_{T}}
&=& -\mathrm{F_x}\sin\theta_0 + \mathrm{F_y}\cos\theta_0 \nonumber\\
&=& - 4\pi\rho_0 g_i^2\int d\theta \sin(\theta-\theta_0)[k^{-}f^{-}(k^{-},\theta) \nonumber\\
&&+ k^{+}f^{+}(k^{+},\theta)],
\label{FT1}
\end{eqnarray}
where $f^{\pm}(k^{\pm},\theta)$ are defined as in Eq. (\ref{FPM}), and $k^{\pm}$ are given by Eq. (\ref{kpm}).
We expand the $f^{\pm}(k^{\pm},\theta)$ in $1/k^{\pm}$ at large $k^{\pm}$, and there are the forms
\begin{eqnarray}
f^{+}(k^{+},\theta) &=& a^{+}_{0} + a^{+}_{1}\frac{1}{k^{+}} + O\Big(\frac{1}{(k^+)^2}\Big),\nonumber\\
f^{-}(k^{-},\theta) &=& a^{-}_{0} + a^{-}_{1}\frac{1}{k^{-}} + O\Big(\frac{1}{(k^-)^2}\Big),
\label{FT2}
\end{eqnarray}
where $a^{\pm}_{0}$ and $a^{\pm}_{1}$ are coefficients need to be calculated.
From Eq. (\ref{FT1}) and Eq. (\ref{FT2}), we have
\begin{eqnarray}
\mathrm{F_{T}} &=& - 4\pi\rho_0 g_i^2 \int d\theta \sin(\theta-\theta_0)\Big\{ k^{-}a^{-}_{0}+k^{+}a^{+}_{0} \nonumber\\
&&+a^{-}_{1} +a^{+}_{1} + O\Big(\frac{1}{v/\lambda}\Big) \Big\}.
\label{FT3}
\end{eqnarray}

Using the expansion of dispersions at large momentum
$\omega^{\pm}_{\mathbf{k}} = \left(\sqrt{(k_x-\lambda)^2+k_y^2}\pm\lambda\right)^2+C^{\pm}(\mathbf{k})m_{\nu}+O(1/k)$,
we obtain
\begin{eqnarray}
f^{+}(k, \theta) &=& \frac{A^{+}k^5 + B^{+}k^4 +O(k^3)}{C^{+}k^5 + D^{+}k^4 +O(k^3)}, \nonumber\\
f^{-}(k, \theta) &=& \frac{A^{-}k^5 + B^{-}k^4 +O(k^3)}{C^{-}k^5 + D^{-}k^4 +O(k^3)},
\label{FT5}
\end{eqnarray}
where the coefficients have
\begin{eqnarray}
&&A^{\pm} = 4\lambda(1\pm\cos\theta),\nonumber\\
&&C^{\pm} = 16\lambda,\nonumber\\
&&B^{+}+B^{-} - \frac{1}{4}(D^{+} + D^{-}) = 16\lambda^2\cos\theta,\nonumber\\
&&D^{+}-D^{-} = 64\lambda^2.
\label{FT6}
\end{eqnarray}
From Eq. (\ref{FT5}) and Eq. (\ref{FT6}), we obtain the coefficients $a^{\pm}_{0}, a^{\pm}_{1}$ as
\begin{eqnarray}
&&a^{\pm}_{0} = \frac{1}{4}(1\pm\cos\theta), \nonumber\\
&&a^{+}_{1}+a^{-}_{1} = 0.
\label{FT7}
\end{eqnarray}

By substituting Eq. (\ref{FT7}) into Eq. (\ref{FT3}),
we find that the coefficient of the term of order $O\Big((v/\lambda)^0\Big)$ is exactly zero.
So the result can be written as
\begin{equation}
\mathrm{F_{T}} = - 4\pi\rho_0 g_i^2\lambda\cdot O\left(\frac{1}{v/\lambda}\right).
\end{equation}
\end{appendix}

\end{document}